\documentclass[hyper]{JHEP3}


\skip\footins = 1\bigskipamount plus 2pt minus 4pt

\usepackage{amsmath,amsfonts,amssymb}
\newcommand{\be}{\begin{equation}}
\newcommand{\ee}{\end{equation}}
\newcommand{\bea}{\begin{eqnarray}}
\newcommand{\eea}{\end{eqnarray}}

\preprint{}

\title{Axion and Right-handed Neutrino in the Minimal SUSY SO(10) Model}

\author{Takeshi Fukuyama, Tatsuru Kikuchi\\
Department of Physics, Ritsumeikan University\\ 
Kusatsu, Shiga, 525-8577 Japan\\
E-mail: \email{fukuyama@se.ritsumei.ac.jp},
\email{rp009979@se.ritsumei.ac.jp}}

\abstract{
The connection between the axion and right-handed neutrinos is explored 
in the framework of the minimal SUSY SO(10) model. The former is related 
to the Peccei-Quinn (PQ) solution to the strong CP problem and the latter is 
to the light Majorana neutrinos through the see-saw mechanism. In this model, 
a relative phase between $({\bf 10,1,3})~(\equiv {\bf \bar{\Delta}}_R)~
\subset {\bf \overline{126}}$ and $({\bf \overline{10},1,3})~
(\equiv {\bf \Delta}_R)~\subset {\bf 126}$ multiplets of 
${\rm SU}(4) \times {\rm SU}(2)_L \times {\rm SU}(2)_R \subset {\rm SO}(10)$ 
becomes a physical degree of freedom identified with the axion. 
Then, the PQ symmetry breaking scale ($\Lambda_{\rm PQ}$) 
and the $B-L$ symmetry breaking scale ($\Lambda_{\rm B-L}$) 
coincide through the VEV of ${\bf \bar{\Delta}}_R$. 
The scalar partner of the lightest right-handed neutrino is regarded 
as the inflaton, which gives a consistent density fluctuation for the CMB. 
}
\begin{document}
\section{Introduction}
The supersymmetric (SUSY) grand unified theory (GUT) has received particular 
attention over the last decade. In particular, with the particle content 
of the minimal supersymmetric standard model (MSSM), the three gauge coupling 
constants converge at the GUT scale $M_{\rm GUT} \simeq 2 \times 10^{16}$ 
[GeV]. In addition to this, recent progress in neutrino physics makes SO(10) 
a plausible candidate for GUTs, since it naturally incorporates 
the see-saw mechanism \cite{see-saw} that can naturally explain 
the lightness of the neutrinos. In particular, minimal SO(10) models are 
very natural to realize since it not only reproduces the low energy 
experimental data but also predicts the unobserved values of absolute 
masses of light neutrinos and heavy right-handed neutrinos and the full 
MNS mixing matrix very restrictively 
\cite{babu, matsudaetal, F-O, type-II, 210-1, Fukuyama:2004xs}. 
Recently well-confirmed atmospheric and solar neutrino oscillation data 
with the see-saw mechanism indicates the scale of the right-handed neutrinos. 
The typical prediction in the minimal SO(10) models 
suggests the scale to be \cite{F-O} 
\be
M_{R1} \simeq 1.2 \times 10^{11} \; {\rm [GeV]}\;,~~
M_{R2} \simeq 1.8 \times 10^{12} \; {\rm [GeV]}\;,~~
M_{R3} \simeq 8.3 \times 10^{12} \; {\rm [GeV]} \;.
\label{MR}
\ee
Even, after this analysis had been performed, there has been remarkable 
progress in the solar neutrino oscillation data from the KamLAND experiment 
\cite{KamLAND}, which does not affect so seriously the above values. 
On the other hand, one of the most likely solutions to the strong CP problem, 
the Peccei-Quinn (PQ) solution, gives us more interesting information at such 
an intermediate scale. Namely, in the PQ solution, we have probably 
an invisible axion with a decay constant $f_{a}$ that is severely 
constrained from astrophysics as: 
\be
10^{9 \pm 1} \; {\rm [GeV]} \lesssim f_a 
\lesssim 10^{12 \pm 1} \; {\rm [GeV]}\;.
\ee
This range is very similar to the scale of the right-handed neutrinos. 
Hence it seems possible that there is some deep connection between 
the two physical scales, the PQ symmetry breaking scale ($\Lambda_{\rm PQ}$) 
and the $B-L$ symmetry breaking scale ($\Lambda_{\rm B-L}$). 
In this paper, we explore the connections of the axion physics 
and the right-handed neutrino physics with the help of the minimal SO(10) 
grand unification model. In SO(10) models, the gauged $B-L$ symmetry can 
play the role of protecting the right-handed Majorana neutrino masses 
from becoming as large as the GUT scale. In this sense, the SO(10) symmetry 
is a necessary gauge symmetry for the GUT to argue about the nature of 
the intermediate energy scale about $10^{13}$ [GeV]. 

In the following, we shall only illustrate the essence of the mechanism 
to connect the axion and right-handed neutrinos. First, the right-handed 
neutrino masses are generated through the following type of Yukawa 
interaction, 
\be
W = Y_{126}^{ij} ~{\bf \bar{\Delta}}_R \,
\nu_{Ri}^T C^{-1} \nu_{Rj} \;.
\ee
This gives the Majorana masses for the right-handed neutrinos, 
$M_R^{ij} = Y_{126}^{ij} \left<{\bf \bar{\Delta}}_R \right>$.  
In general, we can assign a global $U(1)_{PQ}$ charge to these fields. 
For instance, ${\rm PQ}[{\bf \bar{\Delta}}_R] = +2$, ${\rm PQ}[\nu_{R}] = -1$. 
Then after giving rise to the VEV of ${\bf \bar{\Delta}}_R$, 
the global $U(1)_{\rm PQ}$ symmetry would be spontaneously broken and 
there appears a pseudo-NG boson that is later understood as the axion. 
The scalar potential of the ${\bf \bar{\Delta}}_R$ field 
includes the mixing term with the electroweak Higgs doublets, 
\be
V = \lambda~ {\bf \bar{\Delta}}_R {\bf \Delta}_R H_{126} H_{10}\;. 
\ee
Here $H_{10} \equiv ({\bf 1,2,2})$ and $H_{126} \equiv ({\bf 15,2,2})$ 
are the $SU(2)_L$ bi-doublet Higgs fields arising from the ${\bf 10}$ and 
${\bf \overline{126}}$ multiplets of SO(10), respectively, and 
${\bf \Delta}_R$ is required for the anomaly cancellation. 
A linear combination of 
$H_{10}$ and $H_{126}$ Higgs fields becomes the MSSM Higgs doublets 
$H_u$ and $H_d$ that cause the correct electroweak symmetry breaking, 
$SU(2)_L \times U(1)_Y \to U(1)_{\rm em}$. 
This potential would cause a connection between intermediate scale physics 
and the electroweak scale. Since the fields $H_u$ and $H_d$, 
or equivalently $H_{10}$ and $H_{126}$ have couplings to the quarks 
and leptons, a rotation of the Higgs fields $H_{10} \to H_{10} 
\exp(+ 2 i \theta)$ and $H_{126} \to H_{126} \exp(+2 i \theta)$ gives 
a chiral rotation of the quarks and leptons 
$\{q_L, u_R^c, d_R^c, \ell_L, \nu_R^c, e_R^c \} \to \exp(- i\theta) 
\{q_L, u_R^c, d_R^c, \ell_L, \nu_R^c, e_R^c \}$. 
Such a non-trivial transformation indicates an anomalous symmetry 
and it induces an anomalous coupling of the pseudo-NG boson $a(x)$ 
to the gluon field. 
\be
{\cal L}= \frac{a(x)}{f_a} \,
\frac{g_s^2}{32 \pi^2}\, G_{\mu \nu}^A \tilde{G}_A^{\mu \nu}\;, 
\ee
where $g_s$ is the $SU(3)_c$ gauge coupling constant, 
$G_{\mu \nu}^A$ is the gluon field strength and $\tilde{G}^{\mu\nu}_A 
\equiv \frac{1}{2} \epsilon^{\mu \nu \rho \sigma} G_{\rho \sigma}^A$. 
This kind of interaction is used to solve the strong CP problem \cite{PQ}. 
Then the interaction of the axion with the quarks and leptons is given by 
\be
{\cal L} \ =\ \frac{a(x)}{f_a} {\partial}_{\mu} J^{\mu}, 
\ee
where $J^{\mu}$ is a conserved current 
associated with the global $U(1)_{\rm PQ}$ symmetry 
\be
J^{\mu}
= f_a {\partial}^{\mu} a(x) 
+ 2 \sin^2\beta \;\bar{u}_i \gamma^\mu \gamma_5 u_i 
+ 2 \cos^2\beta \;\bar{d}_i \gamma^\mu \gamma_5 d_i 
+ 2 \cos^2\beta \;\bar{e}_i \gamma^\mu \gamma_5 e_i   \;. 
\ee
with $\tan \beta \equiv \left<H_u \right>/\left<H_d \right>$. 
As is usual for the pseudo-NG bosons, the mass of the axion 
is inversely proportional to the decay constant $f_a$ as 
\be
m_a = 0.62  \times 10^{-6} \;[{\rm eV}] \; \times
\frac{10^{13} \;{\rm [GeV]}}{f_a} \;.
\ee
Thus, the right-handed neutrino mass scale suggested from the recent 
neutrino oscillation data implies the appearance of an invisible axion 
with mass
\be
m_a \simeq 7.5 \times 10^{-5} \;{\rm [eV]} \;. 
\ee
Though there have already been many models of the axion \cite{WW, KSVZ, ZDFS}, 
and their applications to GUT models 
\cite{Wise:1981ry, Nilles:1981py, Mohapatra:1982tc, Hall:1995eq}. 
In this paper, we consider now to incorporate the axion into 
the minimal SO(10) model. 

\section{SO(10) model}

In order to realize the axion in the minimal SO(10) model, 
let us denote the right-handed neutrino superfield as $N \equiv \nu_R$. 
As mentioned in the introduction, the masses of the right-handed neutrinos 
are given by an $SU(2)_R$ triplet Higgs field ${\bf \bar{\Delta}}_R$. 
In the minimal SO(10) model 
\cite{babu, matsudaetal, F-O, type-II, 210-1, Fukuyama:2004xs}, 
such a triplet field ${\bf \bar{\Delta}}_R$ can naturally be obtained from 
the ${\bf \bar{\Delta}} = {\bf \overline{126}}$ Higgs field. 
In order to avoid a heavy axion mass, which does not solve the strong CP 
problem, we also impose a discrete symmetry ${\mathbb Z}_3$. 
The corresponding charges with regards to this ${\mathbb Z}_3$ symmetry are 
listed in Table 1. Then, the ${\rm SO}(10) \times {\mathbb Z}_3$ invariant 
superpotential is given by 
\be
W = \Psi_i (Y_{10}^{ij}\; H 
+ Y_{126}^{ij} \;{\bf \bar{\Delta}} ) \Psi_j 
+ m_1 {\bf \overline{\Delta}} {\bf \Delta}
+ m_2 \Phi^2
+ \lambda_1 {\bf \bar{\Delta}} {\bf \Delta} \Phi
+ \lambda_2 {\bf \Delta} H \Phi 
+ \lambda_3 \Phi^3 \;, 
\ee
where $\Psi_i$ is a ${\bf 16}$-dimensional matter multiplet, 
$H$ is a ${\bf 10}$-dimensional multiplet which essentially gives 
a large top Yukawa coupling and $\Phi$ is a ${\bf 210}$-dimensional 
multiplet that is used to break the SO(10) gauge symmetry. 
The details of this potential can be found in 
\cite{Fukuyama:2004xs, Fukuyama:2004ps}. 

The essential point in this framework to generate the PQ axion is as follows: 
the ${\bf \overline{126}}$ and the ${\bf 126}$ are independent fields required 
in order to preserve SUSY, but they always appear in pairs, and the SUSY 
vacuum condition (D-flat condition) can never determine the relative phase 
degree of freedom: 
\be
\left|\left<{\bf \bar{\Delta}}_R \right>\right|^2 - 
\left|\left< {\bf \Delta}_R  \right> \right|^2 =0 \;. 
\ee
This means, the relative phase remains as a physical degree of freedom, 
the so called pseudo-NG boson. Schematically, we can write this fact as 
follows: 
\be
\left< {\bf \bar{\Delta}}_R \right> \sim 
\left< {\bf \Delta}_R \right> \times \exp(i\, \Theta)\;, 
\ee
where the argument field or the pseudo-NG boson $\Theta$ can be 
regarded as the axion. It gives a connection between the $U(1)_{B-L}$ 
symmetry breaking scale $(\Lambda_{B-L})$ and the $U(1)_{\rm PQ}$ 
symmetry breaking scale ($\Lambda_{\rm PQ}$)
\footnote{
Note that since these two fields ${\bf \bar{\Delta}}_R$ and ${\bf \Delta}_R$ 
are completely independent, one of which is used to break the $B-L$ symmetry 
and the other can be used to break the PQ symmetry as well. Remarkably, 
the former symmetry is gauged in SO(10) although the latter one is ungauged, 
hence one of the NG bosons residing in the above fields is absorbed into 
the $B-L$ gauge boson, but the other remains as a physical degree of freedom, 
the axion. 
}.
That is one of our main conclusions in this article. 

After the SO(10) symmetry breaking, we have the following 
superpotential for the matter multiplets: 
\bea
W &=& u_{Ri}^c 
\left(Y_{10}^{ij} H_{10}^u + Y_{126}^{ij} H_{126}^u \right) q_{Lj}
\ +\ d_{Ri}^c 
\left(Y_{10}^{ij} H_{10}^d + Y_{126}^{ij} H_{126}^d \right) q_{Lj}
\nonumber\\
&+& N_{i}^c 
\left(Y_{10}^{ij} H_{10}^u -3\,Y_{126}^{ij} H_{126}^u \right) \ell_{Lj}
\ +\ e_{Ri}^c 
\left(Y_{10}^{ij} H_{10}^d -3\,Y_{126}^{ij} H_{126}^d \right) \ell_{Lj}
\nonumber\\
&+& Y_{126}^{ij} \,N_i^c N_j^c \,{\bf \bar{\Delta}}_R \;.
\eea
Each field can have the PQ charges as listed in Table 1. 
\begin{table}
\begin{center}
\begin{tabular}{|c|c|c|}
\hline \hline
fields & PQ charges & ${\mathbb Z}_3$ charges \\
\hline
$\Psi $ & $-1$ & $\omega^2$ \\
$H$ & $+2$ & $\omega^2$ \\
${\bf \bar{\Delta}}$ & $+2$ & $\omega^2$ \\
${\bf \Delta}$ & $-2$ & $\omega$ \\
$\Phi$ & $0$ & $1$ \\
\hline \hline
\end{tabular}
\caption{PQ and ${\mathbb Z}_3$ charges of the fields ($\omega^3 =1$). }
\end{center}
\end{table}
In addition to this, we have the soft SUSY breaking terms defined as 
follows: 
\be
V_{\rm SOFT} =
m_{\tilde{N}_i}^2 \; |\tilde{N}_i|^2
+ m_{\bar{\Delta}}^2 \; \left|{\bf \bar{\Delta}}_R \right|^2 
+ m_{\Delta}^2 \; \left|{\bf \Delta}_R \right|^2 
+ \left(A_{N}^{ij} \; {\bf \bar{\Delta}}_R \; \tilde{N}_{i} 
\tilde{N}_{j} \ +\ {\rm h.c.} \right) \;,
\ee
where $A_N^{ij}$ is the tri-linear coupling constant which is 
assumed to be proportional to the Yukawa coupling constant $Y_{126}^{ij}$, 
$A_N^{ij} = m_{3/2} Y_{126}^{ij}$. From the superpotential given above, 
we can calculate the scalar potential in the usual way: 
\be
V=
\left|\frac{\partial{W}}{\partial{{\bf \bar{\Delta}}_R}} \right|^2
\ +\ \left|\frac{\partial{W}}{\partial{\tilde{N}}} \right|^2 
\ +\ V_{\rm SOFT} \;,
\ee
that is, 
\be
V=m_{\tilde{N}_i}^2 \, |\tilde{N}_i|^2
+ \left(M_{\rm GUT}^2 + m_{\bar{\Delta}}^2 \right)\, 
\left| {\bf \bar{\Delta}}_R \right|^2
+ \left\{
\left(Y_{126}^{ij} M_{\rm GUT} + A_N^{ij} \right)\, 
{\bf \bar{\Delta}}_R \; \tilde{N}_i^* \tilde{N}_j + {\rm h.c.} \right\} 
 + \cdots \;.
\label{VPQ}
\ee
We regard the scalar partner of the lightest right-handed neutrino 
(sneutrino) $\tilde{N}_1$ as the inflaton, 
that is, we consider the sneutrino inflation scenario \cite{sneutrino}. 
In this case, a condensation of the scalar field 
$\left<\tilde{N}_1 \right>$ causes the inflation 
and the successive reheating processes. 
Then the above potential drives the sneutrino 
$\tilde{N}_1$ (a hybrid inflation \cite{hybrid}) and it determines 
the inflaton (sneutrino) mass to be around 
$m_{\rm inf} \simeq \left(M_{\rm GUT} \, M_{R1} \right)^{1/2} 
\simeq 5.7 \times 10^{13}$ [GeV]. 
The mass scale of the sneutrino as the inflaton is the appropriate one 
for the time of coherent oscillation until the end of inflation 
$H \simeq \Gamma_{\tilde{N}_1}$ ($H$: Hubble parameter), 
namely, it leads to the COBE normalization of 
the primordial density fluctuation \cite{Salopek}
\be
\frac{\delta T}{T}
\ \simeq\ \left(\frac{m_{\rm inf}}{M_P}\right)
\ \simeq\ 10^{-5}\;.
\ee
Here the tree level sneutrino decay rate is given by 
\be
\Gamma_{\tilde{N}_1} 
\ \simeq\ \frac{1}{4 \pi} \left(Y_\nu Y_\nu^\dag \right)^{11} M_{R1}
\ \simeq\ 6.1 \times 10^{7} \; {\rm [GeV]} \;, 
\ee
where $Y_\nu$ is the neutrino Dirac Yukawa coupling matrix, and 
we took the typical value of 
$\left(Y_\nu Y_\nu^\dag \right)^{11} \simeq 4.7 \times 10^{-3}$. 
Thus the reheating temperature in this model is given by \cite{reheat}
\be
T_R \ =\ \left(\frac{45 \;M_{\rm P}^2 }{2 \pi^2 g_*} \right)^{1/4} 
\left(\Gamma_{\tilde{N}_1} \right)^{1/2}
\ \simeq\ 4.0 \times 10^{12} ~{\rm [GeV]} \;.
\ee
After giving rise to the PQ symmetry breaking VEV of the Higgs, 
\be
\left<{\bf \bar{\Delta}}_R \right> \ \simeq\ 
8.3 \times 10^{12} \;{\rm [GeV]} \;,
\ee
the argument of ${\bf \bar{\Delta}}_R$ can be regarded as 
the PQ field or an invisible axion, 
$a(x) \equiv f_a \times \left[\arg \left({\bf \bar{\Delta}}_R \right)
- \arg \left({\bf \Delta}_R \right) \right]$ 
with the decay constant 
$f_a = \left|\left<{\bf \bar{\Delta}}_R \right> 
\right| \ \simeq\ 8.3 \times 10^{12} \;{\rm [GeV]}$. 

Finally, it should be noted that the gauged $B-L$ symmetry included 
in the SO(10) symmetry protects the sneutrinos from having large initial 
values along the existing $B-L$ flat direction. Therefore we can not 
incorporate the simple chaotic inflation scenario \cite{chaotic} 
into the SO(10) models, and we must use the hybrid inflation model. 
Recent WMAP data also supports the fact that multi-field hybrid 
inflation models are preferable to the single field chaotic inflation 
model \cite{WMAP}. 

\section{Conclusion}

It has been found that the Peccei-Quinn solution to the strong CP or 
the axion problem can be embedded in the same framework of the right-handed 
neutrino sector. A complete correspondence between the axion and the Higgs 
that gives a mass for the right-handed neutrino has been obtained based on 
an SO(10) model. The relative phase between ${\bf \bar{\Delta}}_R$ and 
${\bf \Delta}_R$ can be identified with the axion itself. 
The resemblances among the symmetry breaking scale of the PQ symmetry 
and the $B-L$ symmetry or the right-handed neutrino mass scale are thus 
well founded. An axion mass consistent with the neutrino 
oscillation data is found to be $m_a \simeq 7.5 \times 10^{-5}$ [eV]. 
In addition, assuming the sneutrino as the inflaton can naturally be 
embedded into the model with a sneutrino mass around 
$\simeq 10^{13}$ [GeV], 
which is consistent with the density fluctuation of the CMB. 

\acknowledgments
We are grateful to W. Naylor for useful comments. 
The work of T.F and T.K is supported in part 
 by the Grant-in-Aid for Scientific Research 
 from the Ministry of Education, Science and Culture of Japan  
 (\#16540269). 
The work of T.K was supported by the Research Fellowship 
 of the Japan Society for the Promotion of Science (\# 7336).


\begin{thebibliography}{99}
\bibitem{reheat}
This value of the reheating temperature is a little bit high 
in considering the gravitino problem with the hadronic decay \cite{kohri}. 
To avoid such a difficulty, 
we just assume the heavy gravitino $m_{3/2} \gtrsim 100$ [TeV] motivated 
by the anomaly mediated SUSY breaking (AMSB) scenario \cite{anom}.
\end{thebibliography}
\end{document}